\documentclass[11pt,a4paper,fleqn,epsfig]{article}

\usepackage{latexsym}
\usepackage[dvips]{color}
\usepackage{graphicx}
\usepackage{amssymb}
\usepackage{amsmath}
\usepackage{amsfonts}

\usepackage[left=2.0cm,right=2.0cm,top=3.0cm,bottom=3.0cm]{geometry}

\begin{document}

\title{\bf Spectral domain ghost imaging}
\author{Nandan Jha\\
High Pressure \& Synchrotron Radiation Physics Division,
\\Bhabha Atomic Research Centre, Mumbai-400085, India\\
email: nandanj@barc.gov.in}

\date{}
\maketitle

\begin{abstract}

In the last few years,the field of ghost imaging has seen many new developments. From computational ghost imaging to 3D ghost imaging, this field has shown many interesting applications. But the method of obtaining an image in ghost imaging experiments still requires data to be recorded over long duration of time due to averaging over many shots of data. We propose a method to get the intensity correlated images in one shot by averaging over different wavelength components rather than different time components.

\end{abstract}

\vskip 0.50 truecm


\section{Introduction}
Ghost imaging was thought to be a purely Quantum phenomenon \cite{shih1,shih2}, but since pseudo-thermal light ghost imaging was proposed and demonstrated \cite{gatti1,gatti2,shih3}, there has been a great deal of controversy over Quantum/Classical nature of ghost imaging \cite{shih4,shih5,erkmen,shapiro_boyd}. First ghost imaging experiments were performed using entangled photon pairs where one of the entangled photons passes through the object and falls on the bucket detector whereas the other photon is detected by a scanning detector. The image of the object is obtained in coincidence counts of the two detectors. Since this process is inherently dependent on the position entanglement of photon pair to produce the image, ghost imaging was thought to be a quantum phenomenon. On the other hand in thermal light ghost imaging, the light from a pseudo-thermal source is split in two parts using a beam splitter and one part of the beam falls on a bucket detector after passing through the object whereas the other part of the beam directly falls on the scanning detector. The image of the object is obtained when the signals from bucket and scanning detectors are cross-correlated. Since now the image is obtained using a classical source, it seems that the physics behind this process is classical in nature. Indeed the process of thermal light ghost imaging can be explained using classical correlation functions between the fields at scanning and bucket detector plane.

\indent Further developments in this field have meant that many new applications have been found. Thermal ghost imaging \cite{zhang} and lensless pseudo-thermal ghost imaging \cite{cai} have the potential to be applied using simpler tools in many real applications. Computational ghost imaging \cite{shapiro, bromberg} took the field to a different level by removing the need for scanning detector in reference arm. Three dimensional ghost imaging system was demonstrated \cite{padgett} using computational ghost imaging principle. To improve the visibility and contrast to noise ratio, higher order ghost imaging has been proposed \cite{cao,chan} and demonstrated \cite{chen}. Signal to noise ratio and contrast has been improved by other methods such as compressive ghost imaging\cite{katz} and differential ghost imaging\cite{ferri} techniques. Recently there has been progress in using polychromatic sources in ghost imaging to obtain coloured images \cite{padgett2}, to remove the effect of turbulence in the medium \cite{shi} and to improve signal to noise ratio \cite{duan}. These applications and the underlying fundamental questions make this field very interesting. However for ghost imaging to find applications in real world situations, it is important to remove the need for collecting intensity signals for long times - to perform ensemble averaging.

In this article, we consider the case of a broadband light source and show how their large spectral bandwidth can be used to obtain images in a single shot by using the frequency space rather than the temporal space for calculating the correlation between scanning and bucket detector. In Section \ref{Conventional}, we review the theory of conventional ghost imaging for a quasi-monochromatic pseudo-thermal source \cite{erkmen}. In Section \ref{SEGI}, we extend the discussion to broadband light source and find the conditions required to obtain single shot ghost images using spectral domain intensities to perform ensemble averaging. In Section \ref{discussion}, the possible experimental implementation issues are discussed.

\section{Conventional Ghost Imaging}
\label{Conventional}

\begin{figure}[htbp]
\centering
\includegraphics[width=120mm]{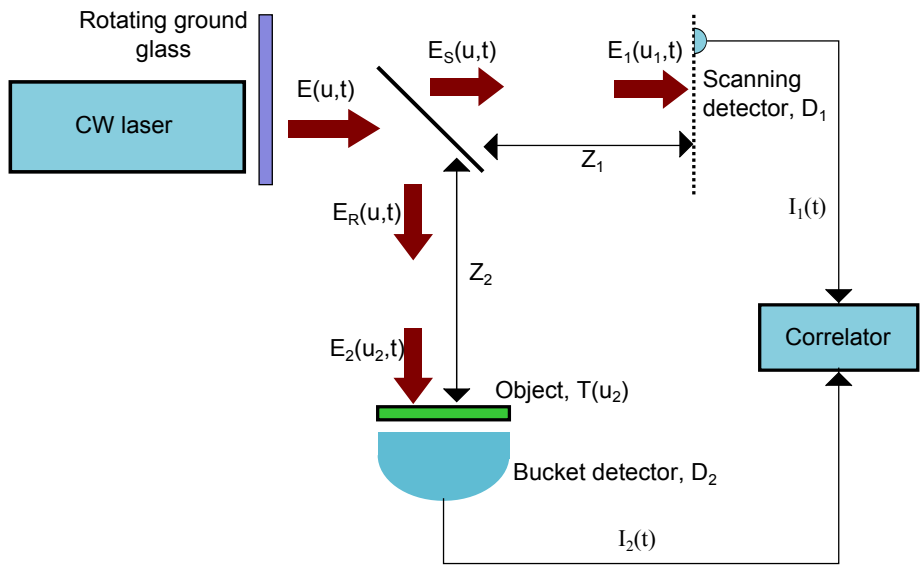}
\caption{Conventional thermal ghost imaging setup.}
\label{fig:setup}
\end{figure}

\indent We consider a simple lens-less pseudo-thermal ghost imaging setup as shown in Fig. \ref{fig:setup}. Light from a pseudo-thermal source is split using a 50:50 beam splitter. One half of the light falls on the object placed at a distance $Z_{2}$ from the source and the light transmitted through the object is collected by a single photon bucket detector $D_{2}$ located behind the object. The other half of light is detected by a scanning detector $D_{1}$ at a distance $Z_{1}$ from the source.

Scanning detector signal for an incident field $E_{1}(u_{1},t)$ is:
\begin{equation}
\label{eq:scanning}
I_{1}\left(u_{1},t\right)=q\eta_{1}\int_{\tau_{1}}^{}\mathrm{d}\tau_{1}\left|E_{1}\left(u_{1},t-\tau_{1}\right)\right|^{2}h_{1}\left(\tau_{1}\right)
\end{equation}
and the bucket detector signal for incident field $E_{2}(u_{2},t)$ is:
\begin{equation}
\label{eq:bucket}
I_{2}(t)=q\eta_{2}\int_{\tau_{2}}^{}\mathrm{d}\tau_{2}\int_{u_{2}}^{}\mathrm{d}u_{2}\left|E_{2}\left(u_{2},t-\tau_{2}\right)\right|^{2}\left|T(u_{2})\right|^{2}h_{2}\left(\tau_{2}\right)
\end{equation}
Here $q$ is the electron charge, $\eta_{l}$ is the quantum efficiency of the detectors, $h_{l}(t)$ represents the finite response time of the photodetector $l$. As can be seen from Eqns. \ref{eq:scanning} and \ref{eq:bucket}, none of the detectors give the object image. But a ``ghost image'' is obtained by correlating the scanning and bucket detector signals.

The ensemble averaged cross correlation function $G^{(2)}(u_{1})$ is estimated by ensemble-averaging the product of photocurrents from the scanning and bucket detectors.
\begin{equation}
G^{(2)}(u_{1}) = \left< I_{1}(u_{1},t)I_{2}(t) \right>_{e}
\end{equation}
If stationary field source is used, then the ensemble averaging can be replaced by a time averaging. Therefore,
\begin{equation}
G^{(2)}(u_{1}) = \left< I_{1}(u_{1},t)I_{2}(t) \right>_{e} = \left< I_{1}(u_{1},t)I_{2}(t) \right>_{t}
\end{equation}

If the source field $E(u,t)$ is a zero-mean, cross-spectrally pure, complex valued Gaussian random process, then the auto-correlation function of the field can be written as:
\begin{equation}
\left<E^{*}(u,t_{1})E(v,t_{2})\right>=K(u,v)R(t_{2}-t_{1})
\end{equation}
with $R(0)=1$ and
\begin{equation}
K(u,v)=\frac{2}{\pi a_{0}^{2}}e^{-\left(\left|u\right|^{2}+\left|v\right|^{2}\right)/a_{0}^{2}-\frac{\left|u-v\right|^{2}}{2u_{0}^{2}}}
\end{equation}
where $a_{0}$ is the beam radius and $u_{0}$ is the spatial coherence radius.

If the source field is divided by a beam splitter as shown in figure \ref{fig:setup}, then
\begin{equation}
E_{S,R}(u,t)=\frac{1}{\sqrt{2}}E(u,t)
\end{equation}
and therefore $E_1$ and $E_2$ also show similar spatial correlations.

The correlation function $G^{(2)}(u_{1})$ can be written as
\begin{equation}
G^{(2)}(u_{1})=\left<I_{1}(u_1)\right>\left<I_2\right>+g^{(2)}(u_1)
\end{equation}
and if the photodetectors have response times much shorter than the field's coherence time, then using the moment-factoring theorem for Gaussian random process,
\begin{equation}
g^{(2)}(u_1)=q^{2}\eta_{1}\eta_{2}\int_{u_{2}}du_{2}\left|\int\int K(x_{1},x_{2})S_{1}^{*}(u_{1},x_{1})S_{2}(u_{2},x_{2})\mathrm{d}x_{1}\mathrm{d}x_{2}\right|^{2}\left|T(u_2)\right|^{2}
\end{equation}
where
$S_{1}$ and $S_{2}$ are the response functions in the scanning and bucket detector arms respectively.
\begin{equation}
\label{eq:responsefunc}
S_{l}(u_{l},x_{l})=-\left(\frac{\iota}{\lambda z_{l}}\right)^{1/2}e^{-\frac{\iota \pi}{\lambda z_{l}}\left(x_{l}-u_{l}\right)^{2}}
\end{equation}
where $l\in (1,2)$. If $z_{1}=z_{2}=z$, then the far field ($\pi a_{0}u_{0}/\lambda z << 1$) expression for $g^{(2)}(u_{1})$ simplifies to
\begin{equation}
\label{eq:g2}
g^{(2)}(u_{1})=q^{2}\eta_{1}\eta_{2}\left(\frac{1}{\pi a_{z}^{2}}\right)^{2}\int \mathrm{d}u_{2}\, e^{-\left|u_{1}-u_{2}\right|^{2}/u_{z}^{2}}\left|T(u_{2})\right|^{2}
\end{equation}
where $a_{z}=z\lambda/\pi u_{0}$ and $u_{z}=z\lambda/\pi a_{0}$ are the far field beam radius and spatial coherence radius of the field at distance z. Similarly,
\begin{equation}
\label{eq:g0}
\left<I_{1}(u_{1})\right>\left<I_{2}\right>=q^{2}\eta_{1}\eta_{2}\left(\frac{1}{\pi a_{z}^{2}}\right)^{2}\int \mathrm{d}u_{2}\, \left|T(u_{2})\right|^{2}
\end{equation}

Eqns. \ref{eq:g2} and \ref{eq:g0} show that by correlating the scanning and bucket detector intensities, a ghost image of object is obtained (Eqn. \ref{eq:g2}) superposed on a constant background term given by Eqn. \ref{eq:g0}. The resolution of the ghost image becomes poorer as we increase $z$ but the field of view increases to $z\lambda/\pi u_0$.

\section{Spectral Ensemble Ghost Imaging}
\label{SEGI}

For a stationary field source, the ensemble average needed to calculate the cross-correlation between scanning and bucket detector signals is replaced by the time averaging process. To remove the requirement of time averaging, some other ensemble is needed over which the cross-correlation can be calculated. In this section, we find the conditions under which the spectral ensemble can be used for the calculation of the correlation between scanning and bucket detector signals.

In a conventional pseudo-thermal ghost imaging experiment, a quasi-monochromatic field is used as described in Section \ref{Conventional}. However, we consider a broadband source field $E(u,\omega)$ which can be written as
\begin{align}
E(u,\omega)&=\sum_{n}a_{n}(u,\omega - \omega_{n})e^{\iota \phi_{n}(\omega - \omega_{n})}\\
&=\sum_{n}E_{n}(u,\omega-\omega_{n})
\end{align}
In a frequency resolved detection setup, if the response time of photodetector is much smaller than the coherence time of each spectral component $E_{n}(u,\omega - \omega_{n})$, then $E_{n}(\omega - \omega_{n})$ $\forall$ $n$ behaves like a quasi-monochromatic source and the analysis of Section \ref{Conventional} is applicable to all $E_{n}$, and $g^{(2)}$ for each spectral component $E_{n}$ can be written as
\begin{equation}
g^{(2)}(u_{1},\omega_{n})=q^{2}\eta_{1}(\omega_{n})\eta_{2}(\omega_{n})\left(\frac{1}{\pi a_{z}^{2}(\omega_{n})}\right)^{2}\int \mathrm{d}u_{2}\, e^{-\left|u_{1}-u_{2}\right|^{2}/u_{z}^{2}(\omega_{n})}\left|T(u_{2})\right|^{2}
\end{equation}
If the following condition is satisfied:
\begin{equation}
\label{eq:condition}
u_{z}^{2}(\omega_{n})=u_{z}^{2}=constant
\end{equation}
then, $g^{(2)}(u_{1},\omega_{n})/C$ is independent of $\omega_{n}$, where
\begin{equation}
\label{eq:calibration}
C = q^{2}\eta_{1}(\omega_{n})\eta_{2}(\omega_{n})\left(\frac{1}{\pi a_{z}^{2}(\omega_{n})}\right)^{2}
\end{equation}
Therefore, the ghost image can be obtained by correlating $I_{1}\times (\pi a_{z}^{2}(\omega_{n})/q\eta_{1}(\omega_{n}))$ and $I_{2}\times(\pi a_{z}^{2}(\omega_{n})/q\eta_{2}(\omega_{n}))$:
\begin{equation}
G_{m}^{(2)}(u_{1},\omega_{n})=\left<\frac{I_{1}I_{2}}{C(\omega_{n})}\right>=\int \mathrm{d}u_{2}\, \left|T(u_{2})\right|^{2} + \int \mathrm{d}u_{2}\, e^{-\left|u_{1}-u_{2}\right|^{2}/u_{z}^{2}}\left|T(u_{2})\right|^{2}
\end{equation}
Condition in Eqn. \ref{eq:condition} can be satisfied if (using Eqn. \ref{eq:responsefunc})
\begin{equation}
\label{eq:lambdacond}
\lambda_{n}z_{1}=\lambda_{n}z_{2}=constant=\alpha \;\; \forall \;\; n
\end{equation}
Eqn. \ref{eq:lambdacond} can be written in its continuous limit ($n\rightarrow \infty$, $\Delta\omega_{n}\rightarrow 0$) as
\begin{align}
z &= \frac{\alpha}{\lambda}\\
&=\frac{\alpha\omega}{2\pi c}
\end{align}
The phase acquired due to travelling distance $z(\omega)$ is $\phi = \omega z(\omega)/c$ which gives
\begin{equation}
\phi = \frac{\alpha}{2\pi c^{2}}\omega^{2}
\end{equation}
which on Taylor expansion about frequency $\omega_{0}$ gives:
\begin{equation}
\phi(\omega)=\frac{\alpha\omega_{0}^{2}}{2\pi c^{2}}+\frac{\alpha \omega_{0}}{\pi c^{2}}(\omega-\omega_{0})+\frac{\alpha}{2\pi c^{2}}(\omega-\omega_{0})^{2}
\end{equation}
Therefore, the condition in Eqn. \ref{eq:condition} essentially requires the incident light to be linearly chirped before impinging on the scanning and bucket detector planes. The required Group Velocity Dispersion (GVD) for a broadband pulse centred at frequency $\omega_{0}$ to be 
\begin{equation}
\label{eq:GVD}
GVD=\frac{\alpha}{2\pi c^{2}}
\end{equation}
Frequency combs will be the ideal candidates which can be used in a ghost imaging experiment of the kind discussed in this section. For a laser pulse at central wavelength 800 nm, $\alpha = 8 \times 10^{-7}$ gives $z(\lambda=800 nm)\approxeq 1\,m$ and from Eqn. \ref{eq:GVD}, the required GVD for such a pulse will be approximately $1.3\times 10^7 fs^2$.

\begin{figure}[htbp]
\centering
\includegraphics[width=120mm]{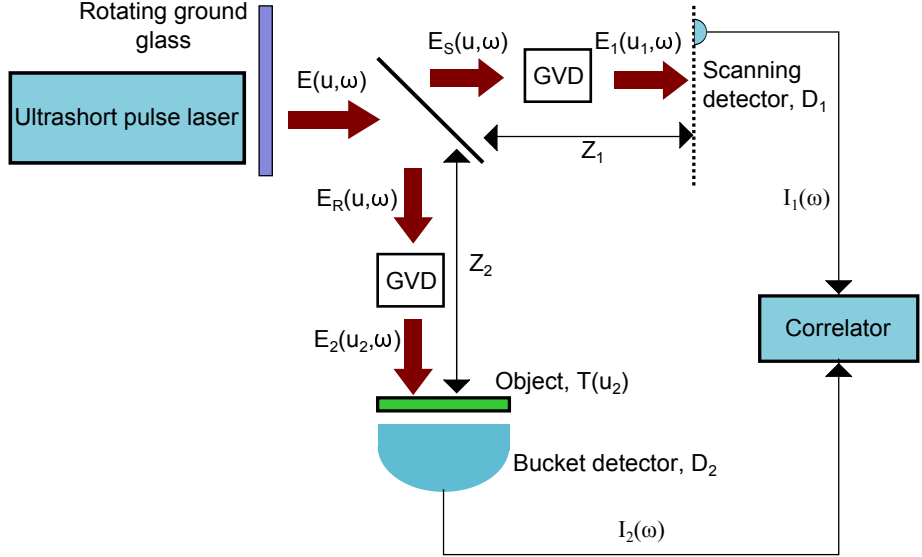}
\caption{Spectral ensemble ghost imaging setup. The bucket and scanning detector are spectrometers and the spectral intensity profile is correlated to obtain the ghost image.}
\label{fig:SEGI_setup}
\end{figure}

If Eqn. \ref{eq:condition} is satisfied, then the ensemble average of product of intensity fluctuations of scanning and bucket detectors are same for all $E_{n}(\omega_{n})$. Hence, the ensemble average over all the spectral components will also give the same ghost image. Therefore, the spectral intensity data can be used (instead of temporal intensity data) as the ensemble over which the averaged cross-correlation can be calculated. In such a case, the ghost image can be obtained in a single shot of data collection (provided enough spectral intensity data points can be obtained in one shot).

\section{Discussion}
\label{discussion}

The experimental set-up required to implement the method outlined in Section \ref{SEGI} is as shown in Fig. \ref{fig:SEGI_setup}. The most important issues involved in such a set-up are:

\noindent (i) Temporal coherence of $E_n(\omega_n)$: The response time of detectors have to be faster than the coherence time of $E_n(\omega_n)$ $\forall$ $n$. This condition implies that the spectrometer used in the scanning and bucket detector plane must have a very fine spectral resolution.

\noindent (ii) It is challenging to add Group Velocity Dispersion (GVD) without any nonlinear dispersion effects. However, the nonlinear dispersions can be minimized by using appropriately designed grating systems \cite{treacy1}-\cite{martinez2}, chirped mirrors \cite{cmirr1}, or by using acoustic phase modulators \cite{tournois}.

\noindent (iii) Exact calibration of the detectors to calculate $C$ (Eqn. \ref{eq:calibration}) as a function of wavelength is required. But such a calibration is required only once for a given source and detector system.

\noindent (iv) Detecting the spectrum at each spatial point in the scanning detector plane ($u_{1}$ plane) requires the spectrometer to be scanned over the whole plane, which means that single shot data cannot be obtained for the whole plane. To overcome this limitation, computational ghost imaging can be employed which will require only one spectrometer in place of a bucket detector.

The technical issues discussed above pose the biggest challenge in experimental implementation of ghost imaging using spectral intensity data. However, if these issues are carefully considered and resolved, such an experiment can be performed with presently available technology.

\renewcommand*{\refname}{\vspace*{-1em}}

\end{document}